\documentclass[12pt,a4paper]{iopart}

\usepackage{graphicx,color}
\usepackage{footnote}
\usepackage{cite}
\usepackage{dcolumn}
\usepackage{epsfig}
\usepackage{hyperref}
\usepackage{iopams}
\usepackage{euscript}
\usepackage{amssymb}
\usepackage{amsthm}
\usepackage{newlfont}
\usepackage{mathtools}
\usepackage{mathrsfs}
\usepackage{subfigure}
\usepackage{changes}
\newcommand{\opunit}{\text{1}\kern-0.22em\text{l}}

\begin{document}

\title[Finite-size scaling of the ferromagnetic Ising model on random regular graphs]{Finite size scaling functions of the phase transition in the ferromagnetic Ising model on random regular graphs}
\author{Suman Kulkarni$^{1,2}$
and Deepak Dhar$^{1}$}
\address{$^1$ Department of Physics, Indian Institute of Science Education and Research Pune,
\\ Homi Bhabha Road, Pashan, Pune 411008, India}
\address{$^2$ Department of Physics and Astronomy, College of Arts and Sciences, University of Pennsylvania, Philadelphia, PA 19104, USA
}

\eads{sumank@sas.upenn.edu, deepak@iiserpune.ac.in}

\begin{abstract}
We discuss the finite-size scaling of the ferromagnetic Ising model on random regular graphs. These graphs are locally tree-like, and in the limit of large graphs, the Bethe approximation gives the exact free energy per site. In the thermodynamic limit, the Ising model on these graphs show a phase transition. This transition is rounded off for finite graphs. We verify the scaling theory prediction that this rounding off is described in terms of the scaling variable $[T/T_c -1] S^{1/2}$ (where $T$ and $T_c$ are the temperature and the critical temperature respectively, and $S$ is the number of sites in the graph), and \emph{not} in terms of a power of the diameter of the graph, which varies as $\log S$. We determine the theoretical scaling functions for the specific heat capacity and the magnetic susceptibility of the absolute value of the magnetization in closed form and compare them to Monte Carlo simulations.
\end{abstract}
\vspace{2pc}
\maketitle
\section{Introduction}\label{sec:intro}
There are only a few problems in science that can be solved exactly and it is important to develop reliable, general approximation methods that are easy to apply. In statistical physics, one widely used approximation method is the mean-field theory. Another method is the Bethe approximation, which goes beyond the simple mean-field treatment by taking into account the correlations between neighboring sites\cite{Bethe1935, Rushbrooke1938}. This method has been applied extensively to a wide range of problems \cite{katsura, peruggi, WH1985, BL1982, ddhar_rltl}. More recent applications of the Bethe approximation use variations like the cluster variation method \cite{cluster} and belief propagation \cite{yedidia}. 
It was shown in Refs. \cite{Kurata,Rushbrooke} that the Bethe approximation becomes exact on an infinite regular tree graph --- i.e. on the Bethe lattice. This is important as it ensures that the approximation does not violate conditions for good behavior, such as the convexity of thermodynamic functions. Also, one avoids problems of the arbitrariness of definition: for example, in the Percus-Yevick approximation for hard spheres, if one uses an approximation for the pair correlation function to calculate the pressure, one gets two different expressions if two different routes are used, whereas both would have been correct if the correlation function were exact \cite{PY1, PY2}.

The Bethe lattice has the disadvantage that it cannot be effectively approximated by any finite lattice. It is well known that the behavior of models on finite tree graphs are very different from the Bethe lattice due to the presence of large number of sites at the boundary \cite{CT1_muller_zittartz,CT2_Eggarter,CT3_Matsuda}. Obtaining the behavior of the Bethe lattice requires determining the behavior in the deep interior of the Cayley tree, away from the boundaries. For a careful derivation of the Bethe approximation for the Ising model, taking into account different possible boundary conditions, see Refs. \cite{Baxter, peruggi}.  

It was later realized that random regular graphs provide a more suitable setting to study the Bethe approximation \cite{rgraphs_BA1, rgraphs_BA2, rgraphs_BA3, rgraphs_BA4, rgraphs_BA5}. Consider a random regular graph of $S$ sites. Let $D$ denote the average length of the shortest loop passing through any randomly picked site. Then, upto a distance of order $D/2$ from any site, the lattice has very few loops. In the limit of large $S$, it can be shown that $D$ diverges and the graph has loops, but {\em locally} resembles the Bethe lattice. The free energy per site for random regular graphs tends to that of the Bethe lattice (with same coordination number) in this thermodynamic limit. Random regular graphs of a given number of sites can also be easily generated and one can numerically study the behavior of various models on them using standard Monte-Carlo methods.

In this paper, we discuss the prototypical ferromagnetic Ising model on random regular graphs. A random regular graph of degree $q$ is a graph drawn at random from the set of all possible graphs of fixed number of sites $S$, where the degree of each vertex in the graph is a fixed number $q$. There is an Ising spin $\sigma_i$ at each site $i$ and the Hamiltonian is the nearest-neighbor ferromagnetic Ising model:
\begin{equation}
H = -J \sum_{<ij>} \sigma_i \sigma_j -h \sum_i \sigma_i,
\end{equation}
where the sum is over all edges $<ij>$ of the graph. 

Let $f_S(\beta, h)$ denote the free energy per site of this model averaged over the ensemble of random regular graphs with $S$ sites. Its limiting value in the thermodynamic limit of large $S$ will be denoted by $f_{\infty}( \beta, h)$. We define $\delta f_S(\beta,h)$ to be the deviation of this average free energy per site from the limiting value
\begin{equation}
    \delta f_S(\beta,h) = f_S(\beta,h) - f_{\infty}(\beta, h)
\end{equation}

From the low and high temperature expansions of $\delta f_S(\beta, h)$, it is easily seen that each term in the expansion varies as $1/S$ for large $S$.  Thus, to leading order in $S$, $\delta f_S(\beta,h) = \phi(\beta,h)/S$, for some function $\phi(\beta, h)$. For any finite $S$, $f_S(\beta, h)$ is an analytic function of $\beta$ and $h$, which develops a singularity in the large $S$ limit.  

Consider a hypercubic lattice in dimensions below the upper critical dimension $4$ having a linear system size of $L$ at a temperature $T = T_c ( 1 + \varepsilon)$. Generally, in critical phenomena, the finite size corrections in the critical region are described in terms of the variable $\varepsilon L^{1/\nu}$, where $\nu$ is the critical exponent that describes the divergence of the correlation length $\xi \sim \varepsilon^{-\nu}$. For $d > 4$, however, this is no longer true.  In terms of the renormalization group theory, this is because the coefficient of the quartic coupling term in the effective Hamiltonian decreases under renormalization, and is hence called `irrelevant'. But, it still affects the amplitudes of the  singularities of various thermodynamic quantities near the critical point. Such variables have been called  `dangerous irrelevant variables'  \cite{fisher1}.  While this  terminology is  generally accepted and used (see for example: Refs. \cite{privman1, cardy1}), these are better termed as coupling constants that `decrease to zero under renormalization, but are still affecting the leading critical behavior' or more simply, as  `vanishing under renormalization but relevant variables' \footnote{What is actually dangerous is not the variable, but an incorrect perception of what is irrelevant.}. In the case of a finite system of $S$ sites in $d > 4$, and for a small $\varepsilon$ and small field $h$, the singular part of the free energy per site $\delta f_S( \varepsilon, h)$ is known to scale \cite{brezin1,rudnick1} as: 
\begin{equation}
\delta f_S(\varepsilon, h) = \frac{1}{S} g( \varepsilon S^{1/2}, h S^{3/4})
\end{equation}
This form of the scaling agrees with the results from simulations of the Ising model on a $5$-dimensional hypercubic lattice \cite{binder1}. Finite-size scaling has also been studied for epidemic type models on scale-free networks in Ref. \cite{karsai}. In this case, the mean-field theory is more complicated because of the large variation in the degree of nodes in the network. The corresponding study for random regular graphs seems to have been not been reported so far.

It is known that the average diameter of random regular graphs of $S$ sites is of the order $\log(S)$ (to leading order in $S$) \cite{diam1, diam2}. Further, the average distance between two randomly picked sites on random regular graphs also  varies as $\log(S)$. For the Ising model on the Bethe lattice, $\xi \sim \varepsilon^{-1/2}$. Assuming that the typical length scale in this problem is the average diameter of the finite graph, we might naively expect the finite-size scaling function to be of the variable $\varepsilon (\log S)^2$. As already mentioned above, this naive expectation is incorrect.  

Differentiating $\delta f_S(\varepsilon, h)$ twice with respect to $\varepsilon$, we see that the quantity $\delta C_v(\varepsilon, h=0)$, which we define as the finite-size deviation of the specific heat capacity per spin from the thermodynamic value has a scaling of the form:
\begin{equation}
\delta C_v( \varepsilon, h=0) = \psi( \varepsilon S^{1/2}, h=0)
\end{equation}

Instead of the standard magnetic susceptibility, we find it convenient to work with the quantity $\chi_{abs} = Var(|M|)/N$. This quantity agrees with the standard definition of the magnetic susceptibility in the low temperature phase  (in the thermodynamic limit). In the high temperature region, it differs by a multiplicative constant. We will show that near $T_c$, the modified magnetic susceptibility at zero field has the scaling form:
\begin{equation}
    \chi_{abs}(\varepsilon, h = 0) = S^{1/2} ~ \omega(\varepsilon S^{1/2})
\end{equation}

The plan of our paper is as follows: In Section \ref{sec:ising_bethe}, we recapitulate the results of the Ising model on the Bethe lattice. We then perform Monte-Carlo simulations of the Ising model on regular random graphs, the details of which are summarized in Section \ref{sec:simulations}, along with the analysis of our results. In Section \ref{sec:theory_scaling}, we discuss the theory explaining  the observed finite-size scaling on these graphs and explicitly calculate the exact scaling functions $\psi$ and $\omega$. 
\section{Ising model on the Bethe lattice}\label{sec:ising_bethe}
A standard approach to obtain the behavior of the Ising model on the Bethe lattice is to study the model on the Cayley tree and only consider the properties of sites deep inside the tree (i.e. far away from the boundary) in the thermodynamic limit \cite{Baxter}.
\begin{figure}[!ht]
    \centering
    \includegraphics[height=6cm, width = 10cm]{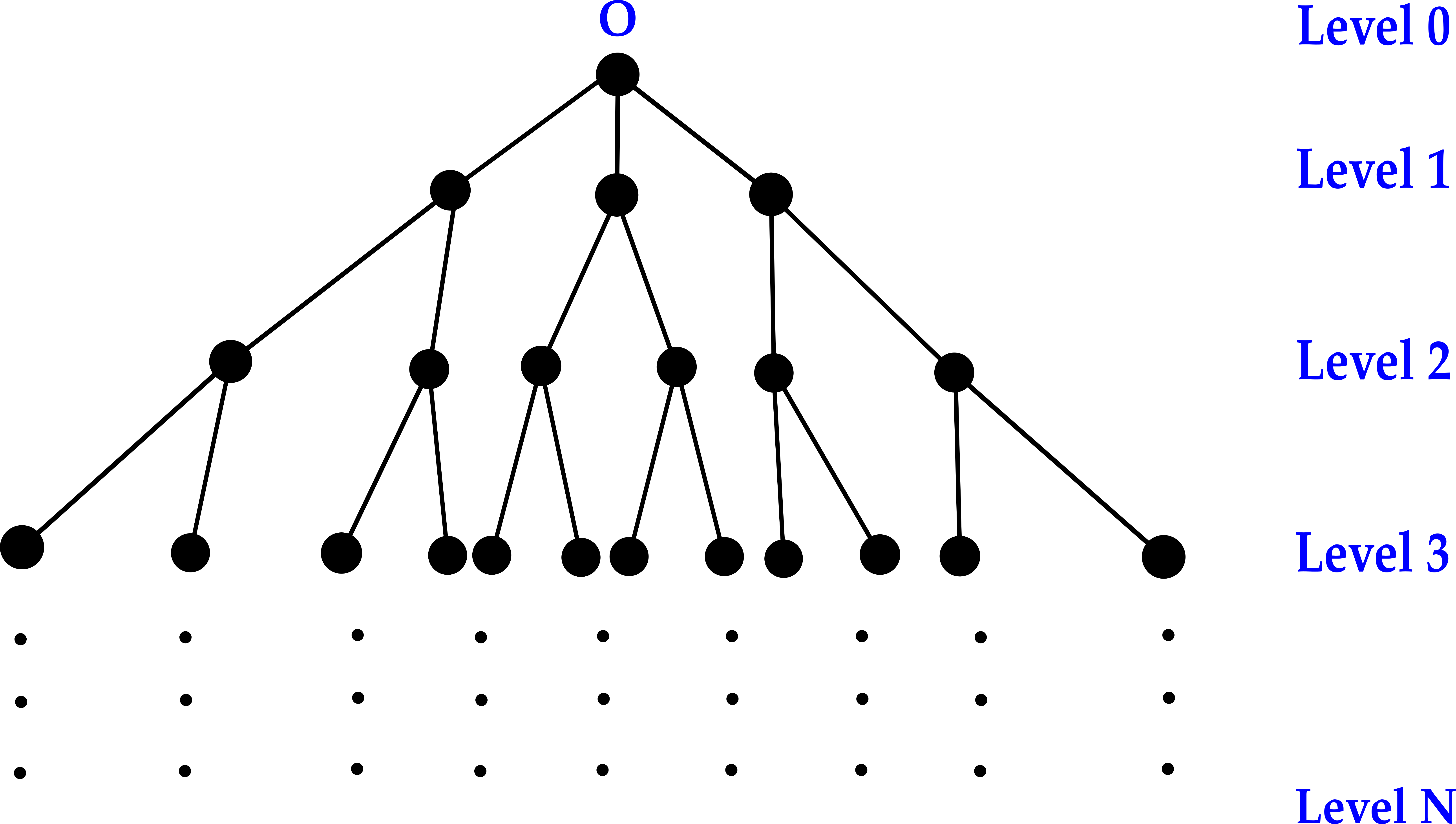}
    \caption{A three-coordinated Cayley tree consisting of $N$ levels centred at site $O$.}
    \label{fig:cayley_tree}
\end{figure}

\begin{figure}
    \centering
    \includegraphics[height=3.5cm, width=7cm]{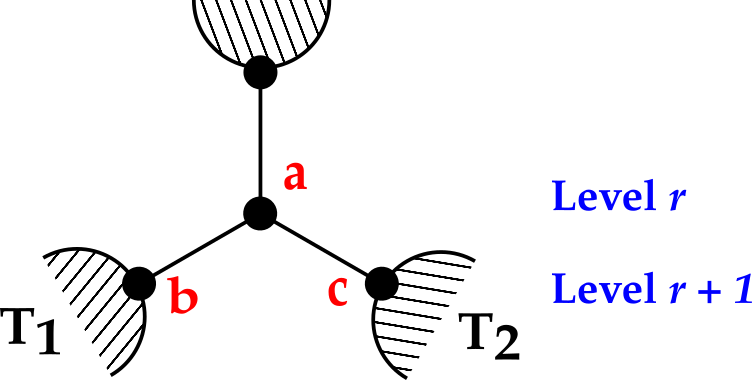}
    \caption{Site $a$ at level $r$ in the Cayley tree connected to two subtrees $T_1$ and $T_2$ via sites $b$ and $c$ at level $r + 1$.}
    \label{fig:cayley_recursion}
\end{figure}

We illustrate this procedure on the simplest case of a three-coordinated tree here. Consider a three-coordinated Cayley tree of $N$ levels centred at site $O$ shown in Fig. \ref{fig:cayley_tree}. Pick an arbitrary site $a$ at level $r$ in the tree. This site is connected to two sites $b$ and $c$ in the  level $r+1$ below. These form the root of two sub-trees $T_1$ and $T_2$ as shown in Fig. \ref{fig:cayley_recursion}.

We define a restricted partition function of a sub-tree rooted at site $\sigma$ as the contribution of the sites ``below" $\sigma$ to the partition function keeping the spin $\sigma$ fixed. The restricted partition function of the sub-tree at level $r$ can then be expressed in terms of the restricted partition functions of the sub-trees $T_1$ and $T_2$ as:  
\begin{equation}
    Z(\sigma_{a}) = \sum_{\sigma_{b}, \sigma_{c}} e^{\beta J(\sigma_{a}\sigma_{b} + \sigma_{a}\sigma_{c})} Z_{T_1}(\sigma_{b}) Z_{T_2}(\sigma_{c})
    \label{pf_two}
\end{equation}
where $\beta = \frac{1}{k_{B}T}$. \\

Each partition function appearing in Eq. (\ref{pf_two}) can be expressed in the form $Z(\sigma_i) = K^{i}(1 + h_{i}\sigma_{i})$, where $h_{i}$ quantifies the effective field at the site due to the sites present in the subsequent levels below it. Using this form, and integrating $\sigma_{b}$ and $\sigma_{c}$ out, the effective field at $\sigma_a$ due to sites below it turns out to be:
\begin{equation}
    \tilde{h} = \frac{x_0(h_{1} + h_{2})}{1 + x_0^{2}h_{1}h_{2}}
    \label{eq:hrec}
\end{equation}
where $x_0 = tanh\beta J$. 

We can repeat the above process at each level in the tree and integrate inwards to get the effective field at a site in the previous level. In other words, Eq. (\ref{eq:hrec}) can be applied recursively at each level. As noted before, we are interested in the behavior of the sites \emph{deep within the lattice}. If we look deep inside the lattice, this effective field  tends to a fixed point value $h^*$ given by
\begin{equation}
    h^{*} = \frac{2x_0h^{*}}{1 + x_0^{2}h^{*2}}
    \label{eq:h*}
\end{equation}

From Eq. (\ref{eq:h*}), we see that there is a critical value of $x_0 = \frac{1}{2}$ at which the behavior of the effective field changes and a phase transition occurs. This corresponds to a $T_{c} = \frac{2J}{k_{B}\; \log (3)}$. In general, the Bethe lattice of coordination number $z$ shows a phase transition at the
critical temperature $T_c$ given by:
\begin{equation}
    \centering
    K_c = \frac{J}{k_{B}T_{c}} = \frac{1}{2} \log \left[\frac{z}{z-2}\right]
    \label{eq:q_bethe}
\end{equation}

For the three-coordinated ($z = 3$) Bethe lattice, the specific heat capacity per site shows a finite jump at the critical temperature. Upon setting $J$ and $k_B$ to 1, it is given by:
\begin{equation}
    C_{v}(T) = \begin{cases}
              \frac{1.5}{(T\cosh(\frac{1}{T}))^2}   &  T \geq T_{c} \\
              \frac{2.25 - 0.75\;\exp(-4/T) - 3\;\exp(-2/T)}{(\sinh(1/T))^4(\cosh(1/T))^2x_0^2(3-\coth(1/T))^2}        &  T < T_{c} \\
             \end{cases}
    \label{eq:cv}
\end{equation}
The magnetic susceptibility per site is:
\begin{equation}
    \chi(T) = \begin{cases}
              \frac{1}{T}\frac{1 + x_0}{(1 - 2x_0)} & T \geq T_{c} \\
              \frac{1}{ T}\frac{(1 + x_0)(1 - x_0)^{3}}{(2x_0 - 1)(3x_0 - 1)^{2}}        &  T < T_{c} \\
             \end{cases}
    \label{eq:chi}
\end{equation}

In systems of finite size, the distribution of the magnetization is double-peaked in the low temperature phase. As a result, the average magnetization of the system is zero even in the ferromagnetic region. To avoid this issue, we find it more convenient to work with the \emph{absolute} magnetization of the system, $|M|$. We use this quantity to calculate the \emph{absolute magnetic susceptibility} $\chi_{abs}$, which is defined in terms of the variance of $|M|$, instead of the variance of $M$ in the standard definition. In the high temperature phase, the distribution of net magnetization is a Gaussian centered at zero and the distribution of $|M|$ is a folded Gaussian distribution. Thus, in this region, the variance of $|M|$ is related to the variance of $M$ by a simple multiplicative factor. In the low temperature region, where the distribution of the magnetization is double-peaked for finite systems, the distribution of $|M|$ will have a single peak and $\chi_{abs}$ gives the correct thermodynamic susceptibility. The quantity $\chi_{abs}$ can hence be written as follows:
\begin{equation}
    \chi_{abs}(T) = \begin{cases}
              \left(1 - \frac{2}{\pi}\right) \frac{1}{T}\frac{1 + x_0}{(1 - 2x_0)} & T \geq T_{c} \\
              \frac{1}{T}\frac{(1 + x_0)(1 - x_0)^{3}}{(2x_0 - 1)(3x_0 - 1)^{2}}        &  T < T_{c} \\
             \end{cases}
    \label{eq:chi_abs}
\end{equation}
As seen, the magnetic susceptibility and the absolute magnetic susceptibility diverge at the critical point.
\section{Simulations}\label{sec:simulations}
We construct a random realization of a regular bipartite graph of $S$ sites (where $S$ is even) with each site having degree $3$ as follows: We first divide the sites into two layers --- called upper and lower layers --- such that each has $S/2$ sites. The sites in each layer are labelled by integers $1$ to $S/2$. Now, we connect site $i$ in upper layer to sites $i$ and $i+1$ in the lower layer. The site $S/2$ in the upper layer is connected to the sites $S/2$ and $1$ in the lower layer. These edges together form a loop of length $S$ in the form of a zig-zag path. Next, we construct a random permutation $P$ on the $S/2$  integers. We then construct an edge between the site $i$ in the upper layer and the site $P[i]$ in the lower layer. Each site in this bipartite graph now has exactly $3$ edges. Our construction is a variation of the rewiring process introduced by Watts and Strogatz for small world networks \cite{watts}. The specific graph we will use is called the Random Locally Tree-like Layered lattice (RLTL lattice) and was introduced in Ref. \cite{ddhar_rltl} to study the Bethe approximation of a system of hard rigid rods. In our case the number of layers is only two.

\begin{figure}[h!]
    \centering
    \includegraphics[height=5.5cm, width=6cm]{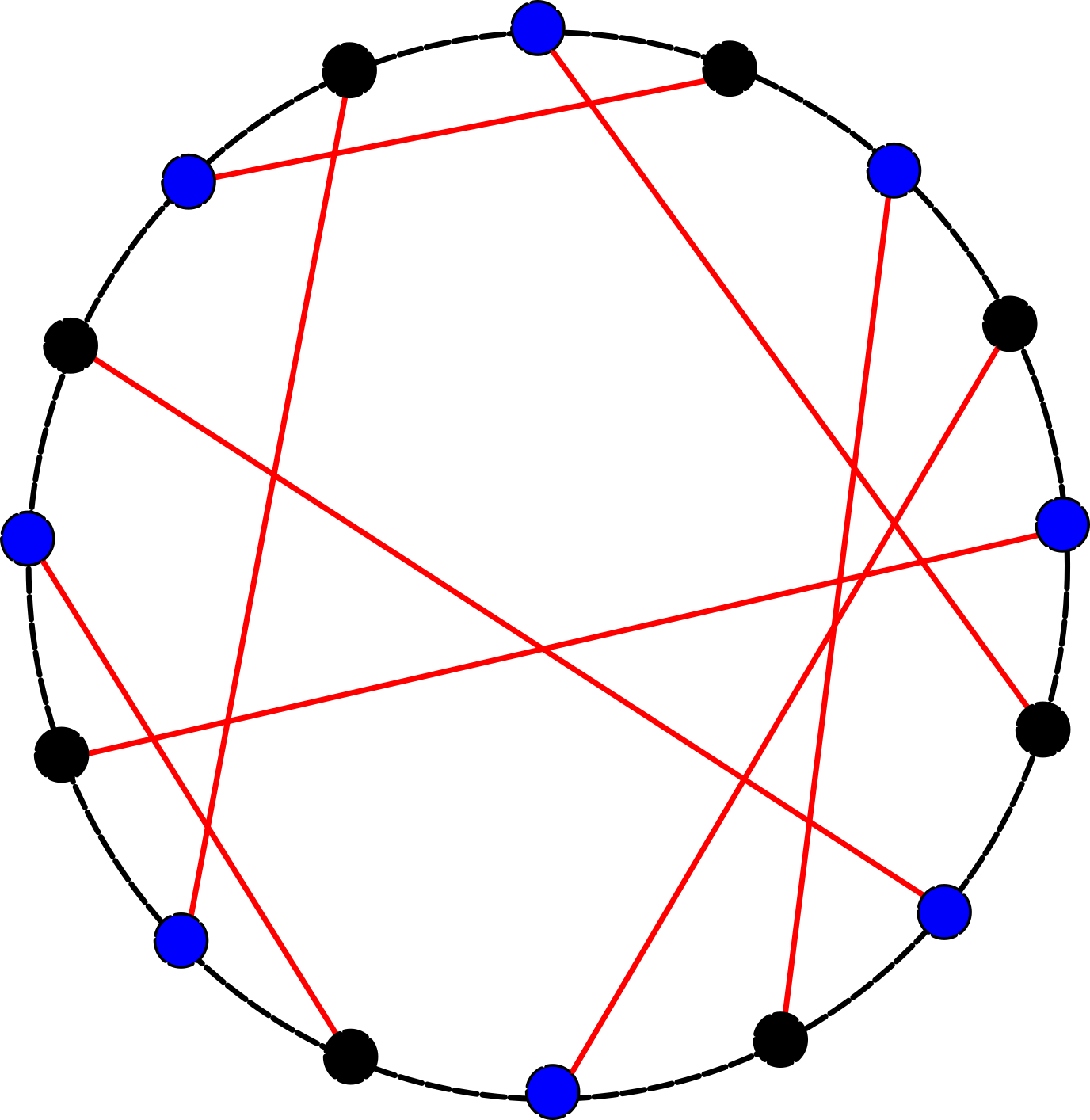}
    \caption{A realization of a three-coordinated RLTL lattice with a single bilayer. The blue and the black nodes denote sites on the upper and lower layer respectively. The black edges denote the zig-zag bonds. The red edges denote the interlayer bonds constructed from a random permutation.}
    \label{fig:rltl_single_bilayer}
\end{figure}

We now define the ferromagnetic Ising model on this graph. There is an Ising spin $\sigma_{\alpha}$  at each vertex $\alpha$, and a ferromagnetic Ising coupling of strength $J$  between two vertices if they are connected by an edge. The Hamiltonian of the system is 
\begin{equation}
H = -J \sum_{<\alpha \beta>} \sigma_{\alpha} \sigma_{\beta}
\end{equation}
where the sum is over all edges $(\alpha, \beta)$ of the graph.

The Ising model simulations were done using the standard  Wolff algorithm, which involves flipping clusters of spins \cite{wolff1}. The basic idea of the algorithm is to build a cluster from a randomly sampled spin. To start the cluster, we pick a spin at random. In the subsequent steps, we look at sites that are connected to the boundary of the cluster and have the same spin. Any such site is added to the cluster with probability $p = 1 - exp(-2 \beta J)$ and will not be a part of the cluster otherwise. Once the growth of the cluster is complete (i.e. no new sites get added to the cluster), all spins in the cluster are flipped. The value of $J$ and $k_B$ is set to one in all our implementations; consequentially, the phase transition occurs at $T = 2/\log(3) \approx 1.82 $. 

We obtained data for $S = 2^{n}$, where n = 7,..., 17. For each lattice size, we take 20 independent realizations of the random graph. We find that there is only a small graph to graph variation in thermodynamic quantities, and this amount of  averaging is adequate.  We run the Wolff algorithm to get a time series of equilibrium configurations at each temperature. To ensure that we are at equilibrium, we discard the first $2 \times 10^3$ samples. We then average over many Wolff clusters, and the number of clusters varied from $5 \times 10^4$ at $T = 1.2$ to $3 \times 10^5$ at $T = 2.2$. For larger sizes ($S \geq 2^{15}$), there is a self-averaging of the different quantities and we find that we can reduce the required number of clusters by half. We report our results after averaging over these 20 independent RLTL realizations (which corresponds to at least $10^{6}$ clusters in total), with error bars denoting the standard error of the estimate.

\begin{figure}[!ht]
    \centering
    \includegraphics[height=6cm, width=7.3cm]{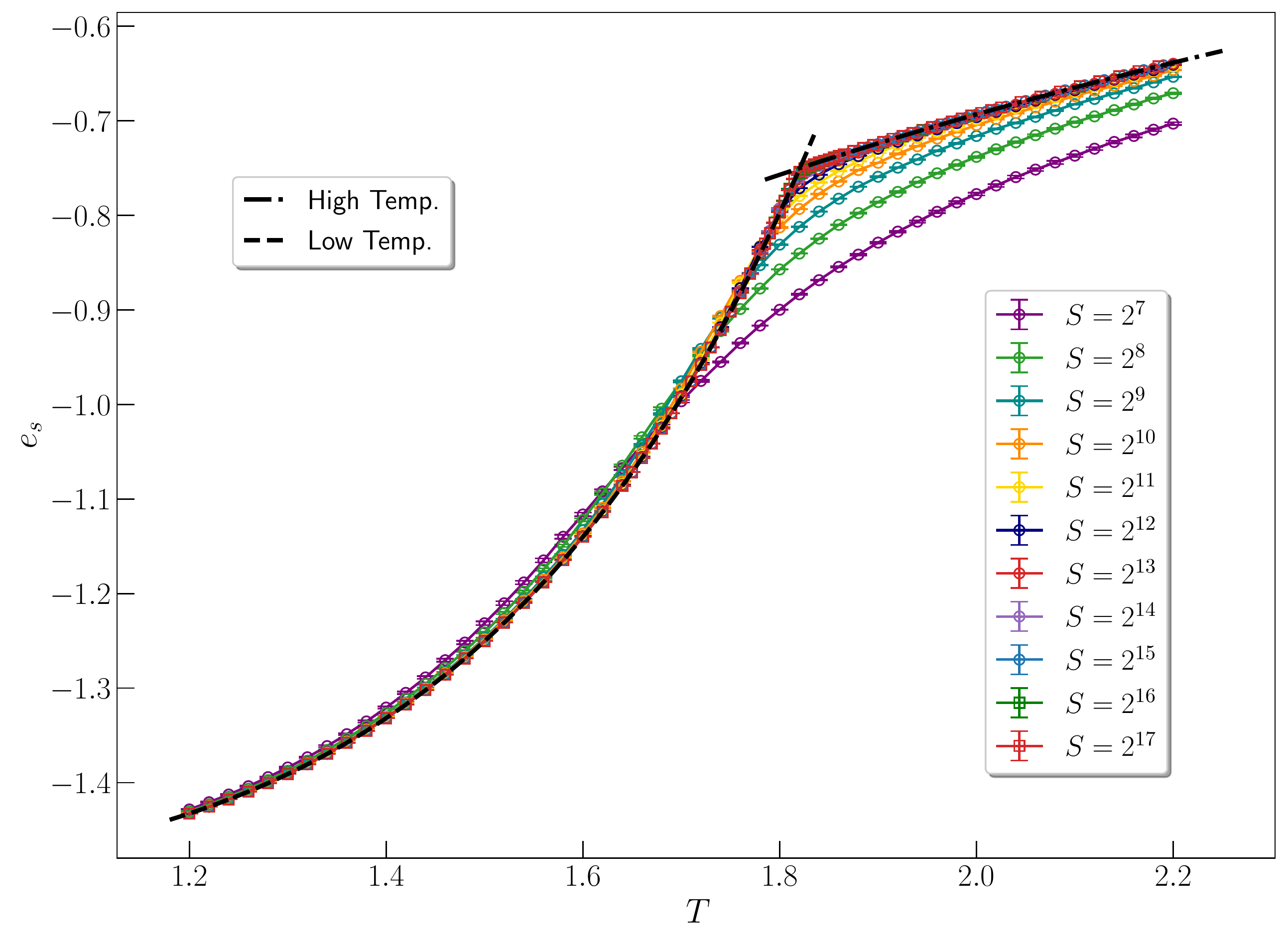}
    \hfill
    \includegraphics[height=6cm, width=7.3cm]{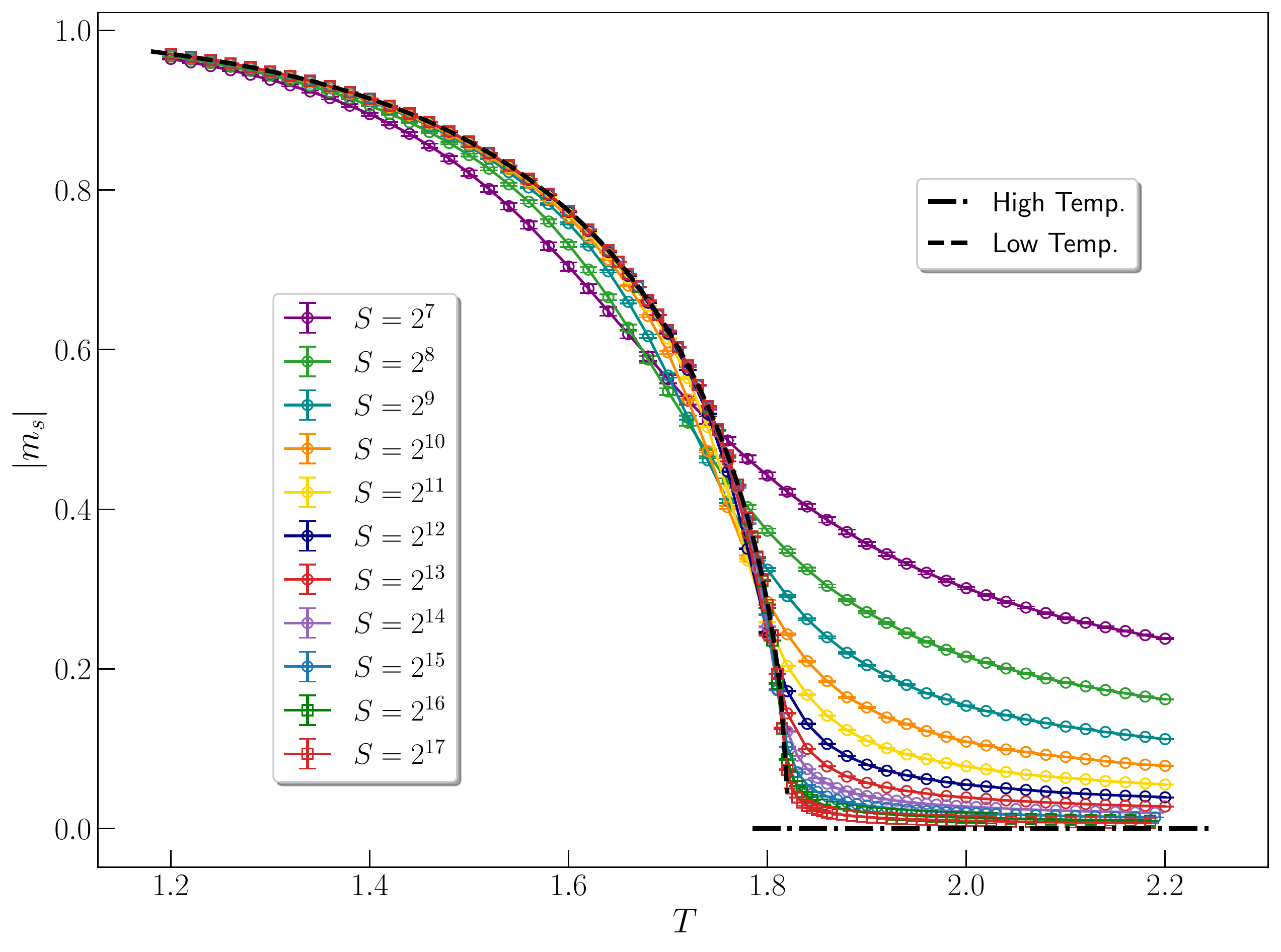}
    \caption{Energy per site ($e_s$) and Absolute magnetization per site ($|m_s|$) as a function of temperature for three-coordinated random regular graphs.}
    \label{fig:RLTL_En_Ms}
\end{figure}

The left panel in Fig. \ref{fig:RLTL_En_Ms} shows the average energy per site ($e_s$) as a function of temperature for lattices of different sizes for the three-coordinated RLTL. The dashed line corresponds to the average energy per site on the Bethe lattice. We see that  the Monte-Carlo simulations for lattice of larger sizes are in good agreement with the value on the Bethe lattice. Lattices with $S = 2^{8}, 2^{9}$ have a maximum error bar of order $10^{-3}$. For larger sizes, the maximum error bars are of order $10^{-4}$ --- primarily in the high temperature region where the size of the Wolff clusters are small. Around the critical point, the errors are of order $10^{-5}$ on these lattices. 

\begin{figure}[!ht]
    \centering
    \includegraphics[height=7cm, width=9.5cm]{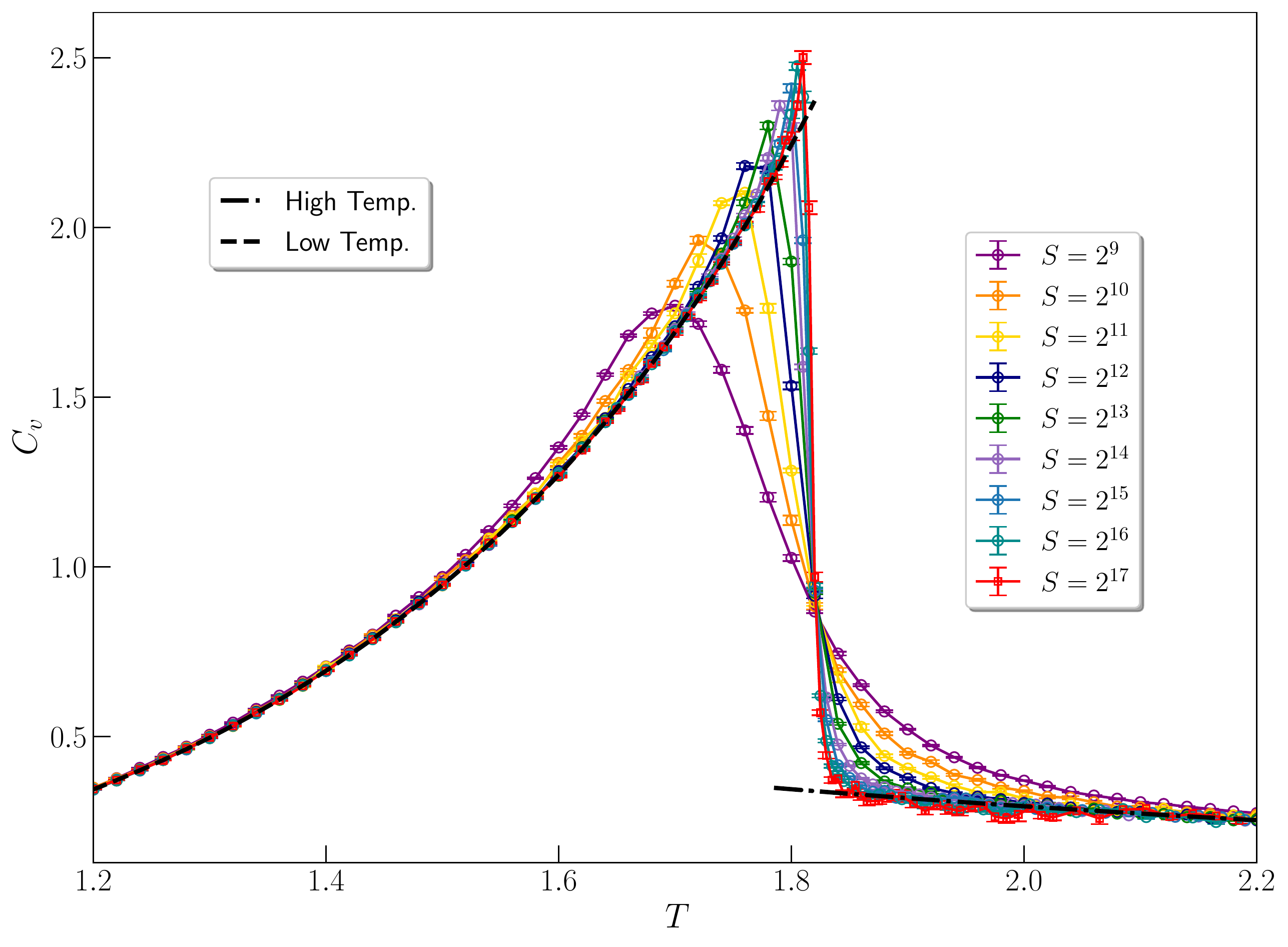}
    \caption{Specific heat capacity per site ($C_v$) as a function of temperature for three-coordinated random regular graphs. The dashed line is the theoretically expected curve (given by Eq. (\ref{eq:cv})).}
    \label{fig:Cv}
\end{figure}

The heat capacity was measured by monitoring the fluctuations in the energy of the system at equilibrium. Fig. \ref{fig:Cv} shows the specific heat capacity per site calculated from our Monte-Carlo data. For lattices of finite size, we do not observe a discontinuity in $C_v$. As we increase the size of the lattice, the plot of $C_{v}$ tends towards the theoretically expected curve. Lattices of size $S < 2^{12}$ had error bars of order $10^{-3}$. For lattices of size $S \geq 2^{12}$, the error at the peak value of $C_{v}$ is $\sim 0.01$ and of order $10^{-3}$ sufficiently away from $T_c$

\begin{figure}[!ht]
    \centering
    \includegraphics[height=7cm, width=9.5cm]{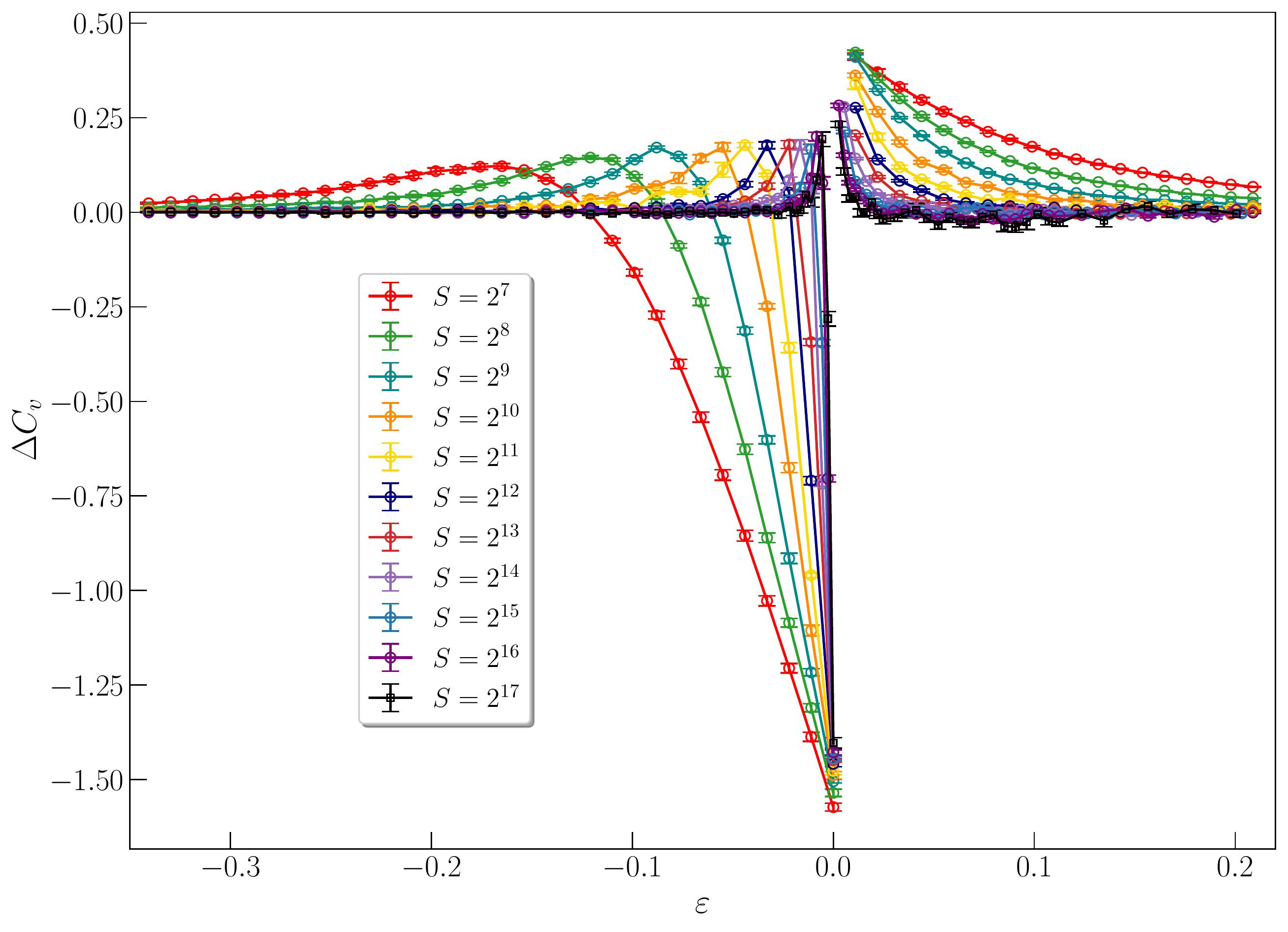}
    \caption{Difference between the specific heat capacity at finite sizes and the specific heat capacity in the thermodynamic limit, $\Delta C_v$, plotted as a function of the reduced temperature for three-coordinated random regular graphs.}
    \label{fig:delta_cv}
\end{figure}

\begin{figure}[!ht]
    \centering
    \includegraphics[height=7cm, width=9.5cm]{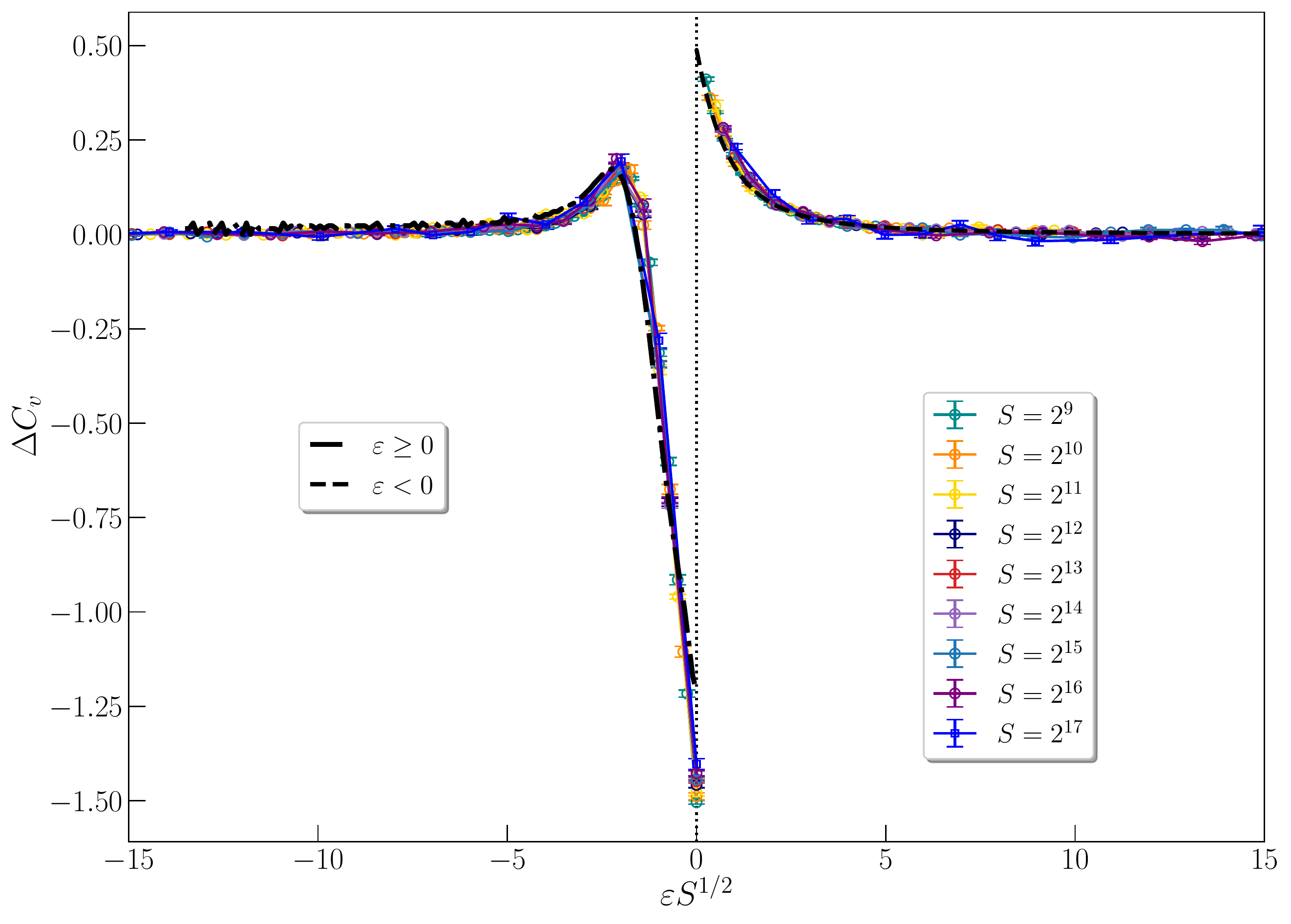}
    \caption{Scaling collapse of $\Delta C_v$. The dashed lines denote the theoretical scaling function given by Eq. (\ref{eq:delta_cv_scaling_theoretical})}
    \label{fig:delta_cv_scaled}
\end{figure}

The specific heat capacity on the Bethe lattice does not diverge at the critical point. To separate the finite-size effects near the critical point for any analytical background, it is useful to consider a quantity $\Delta C_v$, which is defined as:
\begin{equation}
    \Delta C_v(S, T) = C_v (S,T) - C_v (\infty, T).
    \label{eq:delta_cv_definition}
\end{equation}
By construction, this quantity will tend to zero at any fixed $\varepsilon$ for large $S$. Note that for any finite $S$, $C_v(S,T)$ is a smooth function of $\varepsilon$.  This implies that $\Delta C_v(S, T)$ has a discontinuity at $\varepsilon =0$, which comes from the discontinuity in $C_v (\infty, T)$.  A plot of $\Delta C_v$ for lattices of different sizes using Monte-Carlo data is shown in Fig. \ref{fig:delta_cv}.  In Fig. \ref{fig:delta_cv_scaled}, we show that curves for different values of $S$ fall onto a single scaling curve if we plot $\Delta C_v$ as a function of the single scaling variable $\varepsilon S^{1/2}$.

We also study the absolute value of net magnetization and its variance. The right panel of Fig. \ref{fig:RLTL_En_Ms} shows the averaged absolute magnetization per site ($m_s$) as a function of temperature from our Monte-Carlo studies, with the theoretical expression shown using dashed lines. For lattices with $S \leq 2^{10}$, the maximum error in $|m|$ occurs around the critical point and is of order $10^{-3}$ . Lattices with $S > 2^{10}$ had a maximum error of order $10^{-4}$ around the critical point. 

\begin{figure}[h!]
    \centering
    \includegraphics[height=7cm, width=9.5cm]{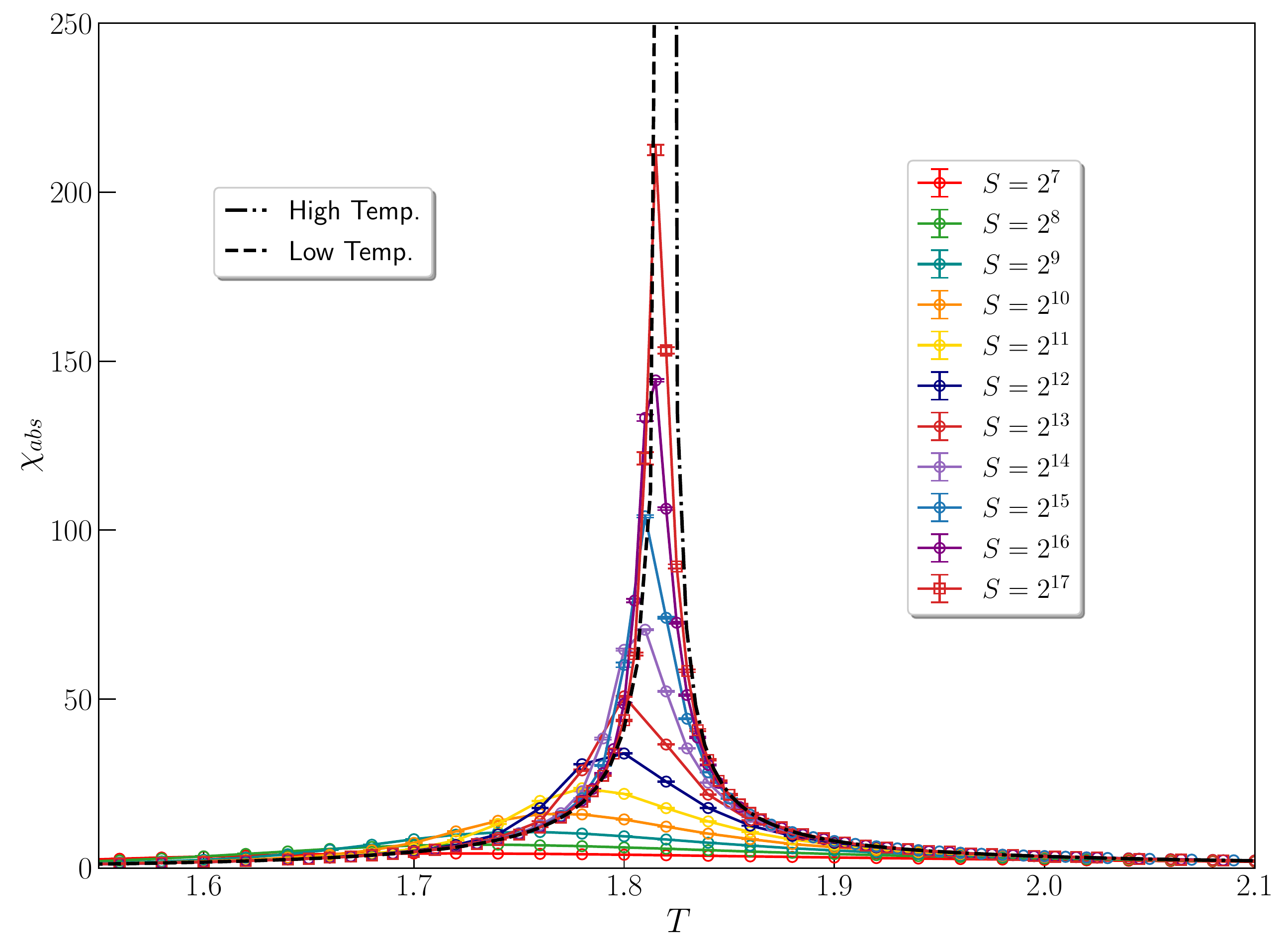}
    \caption{Magnetic susceptibility per site calculated using the absolute value of the magnetization ($X_{abs}$) as a function of temperature for three-coordinated random regular graphs. The dashed line is the theoretically expected curve (given by Eq. (\ref{eq:chi})).}
    \label{fig:X}
\end{figure}

\begin{figure}[!ht]
    \centering
    \includegraphics[height=7cm, width=9.5cm]{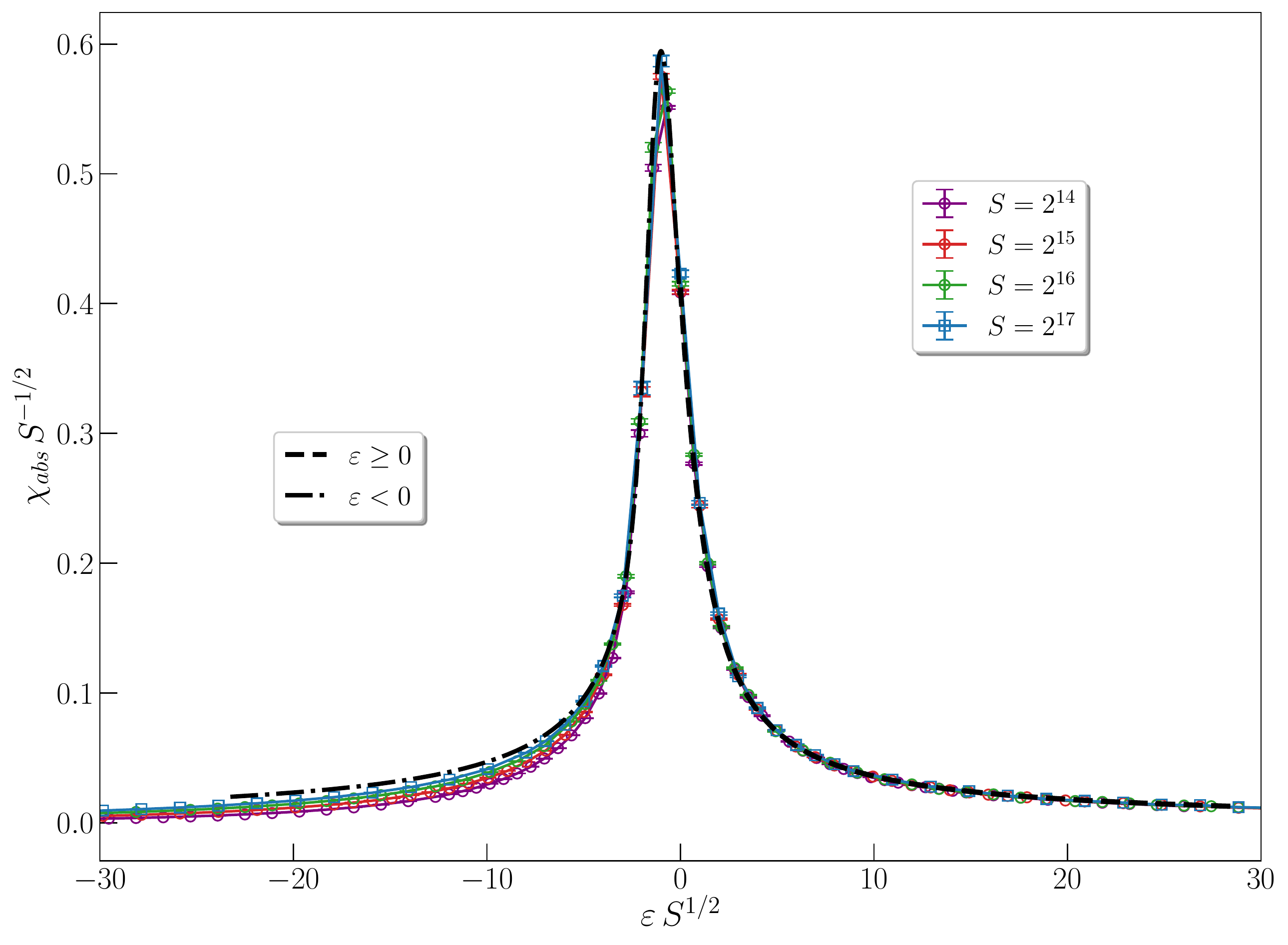}
    \caption{Scaling collapse of the plot of $\chi_{abs}$. The dashed lines denote the theoretical scaling function given by Eq. (\ref{eq:x_abs_scaling})}.
    \label{fig:x_scaled}
\end{figure}

The plot of $\chi_{abs} $ as a function of temperature for different values of $S$ is shown in Fig. \ref{fig:X}. Across all lattice sizes, the maximum error in $\chi_{abs}$ occurs at the peak and the percentage error at this data point is of order $0.1\%$. Since we are working with lattices of finite size, $\chi_{abs}$ does not diverge at $T_{c}$, but its peak value increases with $S$. We were able to obtain a collapse when $\chi_{abs} S^{-1/2}$ is plotted versus $\varepsilon S^{1/2}$. This is shown in Fig. \ref{fig:x_scaled}. 

\section{Theoretical derivation of the scaling functions}\label{sec:theory_scaling}

\begin{figure}[!ht]
    \centering
    \includegraphics[height=7cm, width=16cm]{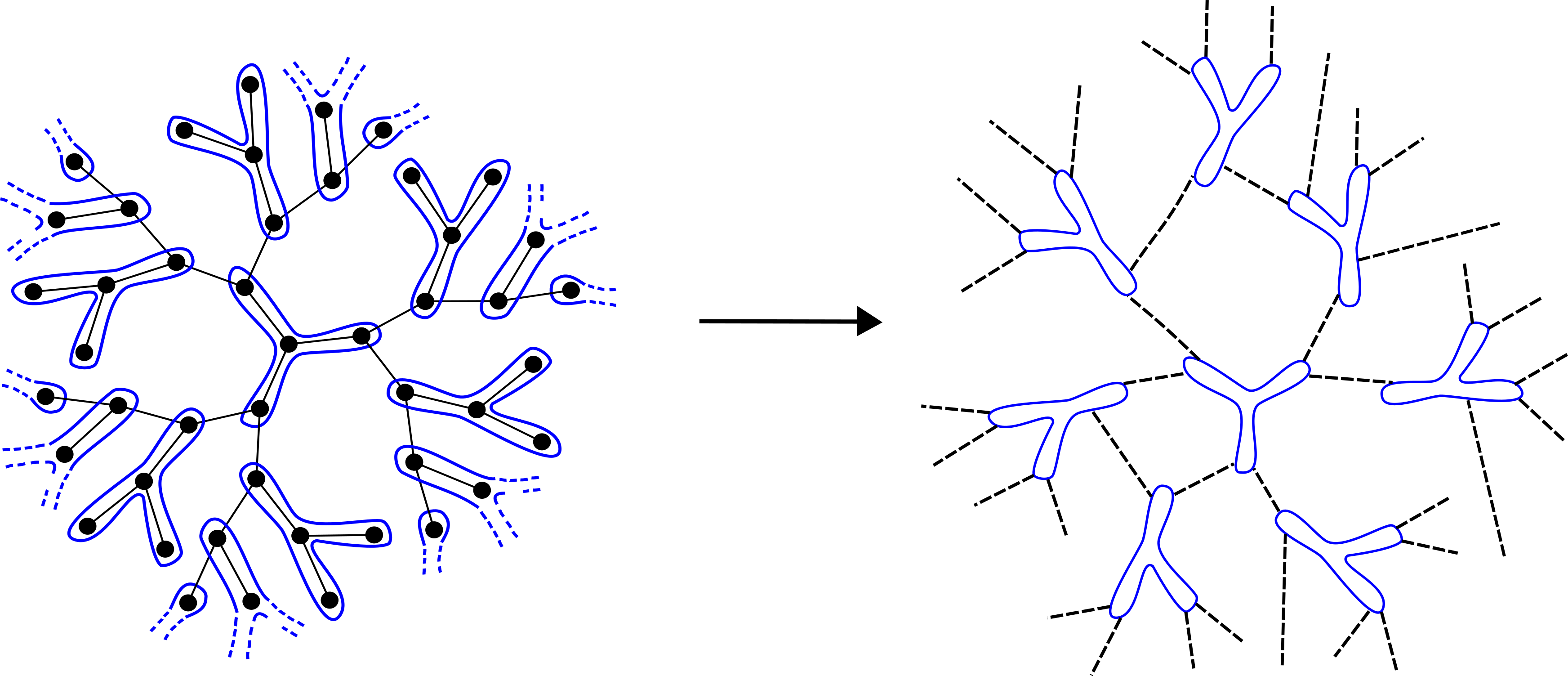}
    \caption{Maximal covering of blocks of spins of size four on the random regular graph. This figure shows a portion of the lattice.}
    \label{fig:renorm1}
\end{figure}

The key concept underlying the scaling theory is that of coarse-graining. However, coarse-graining on random regular graphs is not straightforward to define. Suppose we try to divide the sites of the lattice into non-overlapping `compact' blocks,  say of four spins each, with each block consisting of a central site and its three nearest neighbors (Fig. \ref{fig:renorm1}). It is not clear that  we can cover all sites this way, for a given choice of the shape of the block. We can, however, construct a maximal cover with some sites remaining uncovered. We then define a block spin $S_B$ for each block $B$. The renormalized Hamiltonian can be obtained by taking the log of the marginal probability distribution Prob$(\{ S_B \})$, which is obtained by integrating over all the original spin variables. 

From Figure \ref{fig:renorm1}, it is clear that the structure of the graph formed by blocks would also be locally tree-like, up to some distance. But the coordination number of this graph is higher: for any given block, there are six blocks that have some direct link to some spin interacting directly with a given block. 

As the number of neighbors of a block increases very fast under any such a coarse-graining, it seems plausible that after a small number of steps of renormalization (of ${\mathcal O}[\log \log S]$), the effective Hamiltonian of the system tends to the Hamiltonian in which each block  spin is coupled to every other block with equal strength. This is also seen from the fact that the distribution of distances between vertices on the random graph is very sharply peaked, and most vertices are at some distance $R^*$, or $R^* \pm 1$. Thus, each spin interacts with most other spins with nearly equal strength. We  denote this Hamiltonian by ${\cal H}_{MF}$:
\begin{equation}
{\cal H}_{MF} = -\frac{J_{eff} }{2 S} (\sum _i S_i)^2
\end{equation}

While ${\cal H}_{MF}$ is clearly not the Hamiltonian in our problem, based on the previous arguments, we expect that the finite-size scaling functions for both $H$ and ${\cal H}_{MF}$ are the same. The effective coupling $J_{eff}$ is a function of the starting value of $\beta J$ in the renormalization procedure, and the system is in the ferro or paramagnetic phase depending on whether $ J_{eff}$ is greater or less than $1$. At the critical point $\beta^*$, where $tanh(\beta^* J) = 1/2$, $J_{eff}$ has to be equal to 1. We are unable to determine $ d J_{eff}/ d(\beta J)$ at $\beta^* J$ from our arguments as we are not able to carry out the renormalization procedure exactly. The scaling theory involves its  value, and   it will be treated as an undetermined parameter in our discussion. 

For the Hamiltonian ${\cal H}_{MF} $, it is straightforward to write the marginal distribution of the total magnetization $M =\sum_i S_i$. It is given by:
\begin{equation}
    {\mathrm Prob}(M) = \frac{1}{Z_S(\beta)} \binom{S}{\frac{S+M}{2}} e^{\beta J_{eff} M^2/(2S)}
    \label{part_fun}
\end{equation}
where 
\begin{equation}
    Z_{S}(\beta) = \sum_{M} \left[ \binom{S}{\frac{S+M}{2}} e^{\beta J_{eff}
    M^2/(2S)}\right]
    \label{part_fun}
\end{equation}
The sum over $M$ in the equation above can be replaced by an integral. We also switch variables, changing from the total magnetization $M$ to the magnetization per site $m = M/S$. The partition function now looks as follows:
\begin{equation}
Z_{S}(\beta) = S \int_{-1}^{+1} dm \exp \left[ -S F(m)\right]
\label{part_fun_integral}
\end{equation} 

where 
\begin{equation}
F(m) = -\beta J_{eff} m^2 +\left[\frac{1+m}{2} \log \frac{1+m}{2}  + \frac{1-m}{2} \log \frac{1-m}{2}\right]
\label{free_energy}
\end{equation}
Note that above the critical temperature, the system is in the paramagnetic phase where $m$ is zero. Sufficiently close to the critical point (i.e. for small $\varepsilon$), $m$ will still be small, and one can Taylor expand the free energy per site in Eq. (\ref{free_energy}) around $m = 0$. This leads to a free energy of the following form, which is familiar to us from the Landau theory of phase transitions \cite{landau1, landau2}.
\begin{equation}
F(m) = a\varepsilon m^2 + b m^4
\end{equation}
This equation is the same as that postulated in \cite{brezin1, rudnick1}. Here, $a$ and $b$ are constants that only depend on the coordination number of the original graph. Define the function
\begin{equation}
\phi(u) = \int_{-\infty}^{+\infty} dx \exp\left[\frac{-u x^2}{2} - \frac{x^4}{4}\right]
\end{equation}
We note that $\phi(u)$ can be expressed in terms of Bessel functions of fractional order \cite{abramowitz} in the region $u \ge 0$
\begin{equation}
\phi(u) = \frac{1}{\sqrt{2}} e^{\frac{u^2}{8}} \sqrt{u} \; K_{\frac{1}{4}}\left(\frac{u^2}{8}\right)\text{ if } u > 0
\end{equation}
In Eq. (\ref{part_fun_integral}), we change variables to $ x = m ( 4 b S)^{1/4}$, and approximate the limits of integration from $-\infty$ to $+\infty$, which gives: 
\begin{equation}
Z_S(\beta) = \left(\frac{S^3}{4b}\right)^{1/4} \phi\left[\frac{a}{b^{1/2}} \varepsilon S^{1/2}\right] \text{ if } \varepsilon > 0
\label{eq:part_fun_1}
\end{equation}
Let us define $\tilde{\varepsilon} = \frac{a}{b^{1/2}} ~ \varepsilon S^{1/2}$. We can determine $C_v(\varepsilon, h=0)$ by taking two derivatives with respect to $\tilde{\varepsilon}$. This gives:
\begin{equation}
    C_v =  \frac{a^2}{b} ~ \frac{d^2}{d \tilde{{\varepsilon}^2}} \log \left[\phi( \tilde{\varepsilon})\right]
\end{equation}
$\delta C_v$ can be found by subtracting the value of $C_v$ in the thermodynamic limit:

\begin{align}
\delta C_v  =& ~ \frac{a^2}{b} ~ \frac{d^2}{d \tilde{\varepsilon}^2}  \log \left[\phi( \tilde{\varepsilon})\right] - K \theta( -\varepsilon)\\
            =& ~ \psi(\varepsilon S^{1/2})
\label{eq:delta_cv_scaling_theoretical}
\end{align}

where $K = 3/2$ and $\theta(x)$ is the unit step function, which comes from subtracting the term corresponding to the thermodynamic limit.

We can similarly write $<M^2> $ in terms of $d[\log(\phi(\tilde{\varepsilon}))]/d\tilde{\varepsilon}$.  The quantity $<|M|> $ can be expressed in terms of the error 
function. This gives us an expression for the absolute magnetic susceptibility:
\begin{align}
\chi_{abs} = & ~ Var (|M|) \nonumber \\
           = & ~ \frac{S^{1/2}}{2b^{1/2}T_{c}} \left\{ \frac{1}{2} ~ \tilde{\varepsilon} ~ \left(\frac{K_{\frac{3}{4}}\left(\frac{\tilde{\varepsilon}^2}{8}\right)}{K_{\frac{1}{4}}\left(\frac{\tilde{\varepsilon}^2}{8}\right)}-1\right) - \frac{2 \pi ~ e^{\frac{\tilde{\varepsilon}^2}{4}} \; \left(\text{erfc}\left(\frac{\tilde{\varepsilon}}{2}\right)\right)^2}{\tilde{\varepsilon} \left(K_{\frac{1}{4}}\left(\frac{\tilde{\varepsilon}^2}{8}\right)\right)^2} \right\} \\
           = & ~ S^{1/2} ~ \omega(\varepsilon S^{1/2})
\label{eq:x_abs_scaling}
\end{align}
We have determined the form of the scaling functions upto the non-universal scaling constants $a$ and $b$. The values of $a$ and $b$ in our case can be determined by fitting our theoretical scaling functions to the Monte-Carlo data. For the region $\varepsilon \ge 0$, the exact scaling functions for $X_{abs}$ and $C_v$ have been obtained in terms of known functions, while for $\varepsilon < 0$, we can obtain them numerically using the partition function in Eq. (\ref{eq:part_fun_1}). The re-scaling was done separately for $\varepsilon \ge 0$ and $\varepsilon < 0$. Our best fit values for $a$ and $b$ using this procedure are $(0.289, 0.019)$ for $\varepsilon < 0$ , and $(0.275, 0.018)$ for $\varepsilon \ge 0$.

Similarly, for $\Delta C_v$, the X-axis needs to be scaled by a factor $\frac{a}{b^{1/2}}$, while the double derivative of $log(\phi)$ term needs to be multiplied by $\frac{a^2}{b}$. Observe that the scale factors are related to each other; the latter being the square of the former. To obtain the fit seen in Fig. \ref{fig:delta_cv_scaled}, the scale factors ($\frac{a}{b^{1/2}}$, $\frac{a^2}{b})$ used were (1.9, 3.62) for $\varepsilon \ge 0$ and (1.85, 3.03) was $\varepsilon < 0$. While the fit for $\varepsilon \ge 0$ is consistent with our expectation, we see that the fit in the region $\varepsilon < 0$ shows some deviation. This is likely due to numerical errors in calculating the double derivatives.
\subsection{Corrections to scaling}
While we are able to obtain the asymptotic scaling function from this theory, we see in Fig. \ref{fig:x_scaled} that there are significant corrections to the theoretical scaling function for $\varepsilon S^{1/2} < -5$ (this corresponds to $\varepsilon \sim -0.02)$. We find that these corrections can be explained by adding an $\varepsilon$ dependence to the coefficient of $M^4$ in the free energy as described in Eq. (\ref{eq:correction_to_scaling}). This correction is formally a ``correction to scaling" term in the scaling theory.
\begin{equation}
    F(M) = \frac{a \varepsilon M^2}{S} + \left( b + d\varepsilon \right) \frac{M^4}{S^3}
    \label{eq:correction_to_scaling}
\end{equation}
where $d$ is a constant. Casting this equation in terms of the re-scaled variables $\tilde{M} = M/S^{3/4}$ and $\tilde{\varepsilon} = \varepsilon/S^{1/2}$, we obtain:
\begin{equation}
    F(\tilde{M}) = a \tilde{\varepsilon} \tilde{M}^2 + \left( b + \frac{d\tilde{\varepsilon}}{S^{1/2}} \right) \tilde{M^4}
\end{equation}

\begin{figure}[!ht]
    \centering
    \includegraphics[height=7cm, width= 9cm]{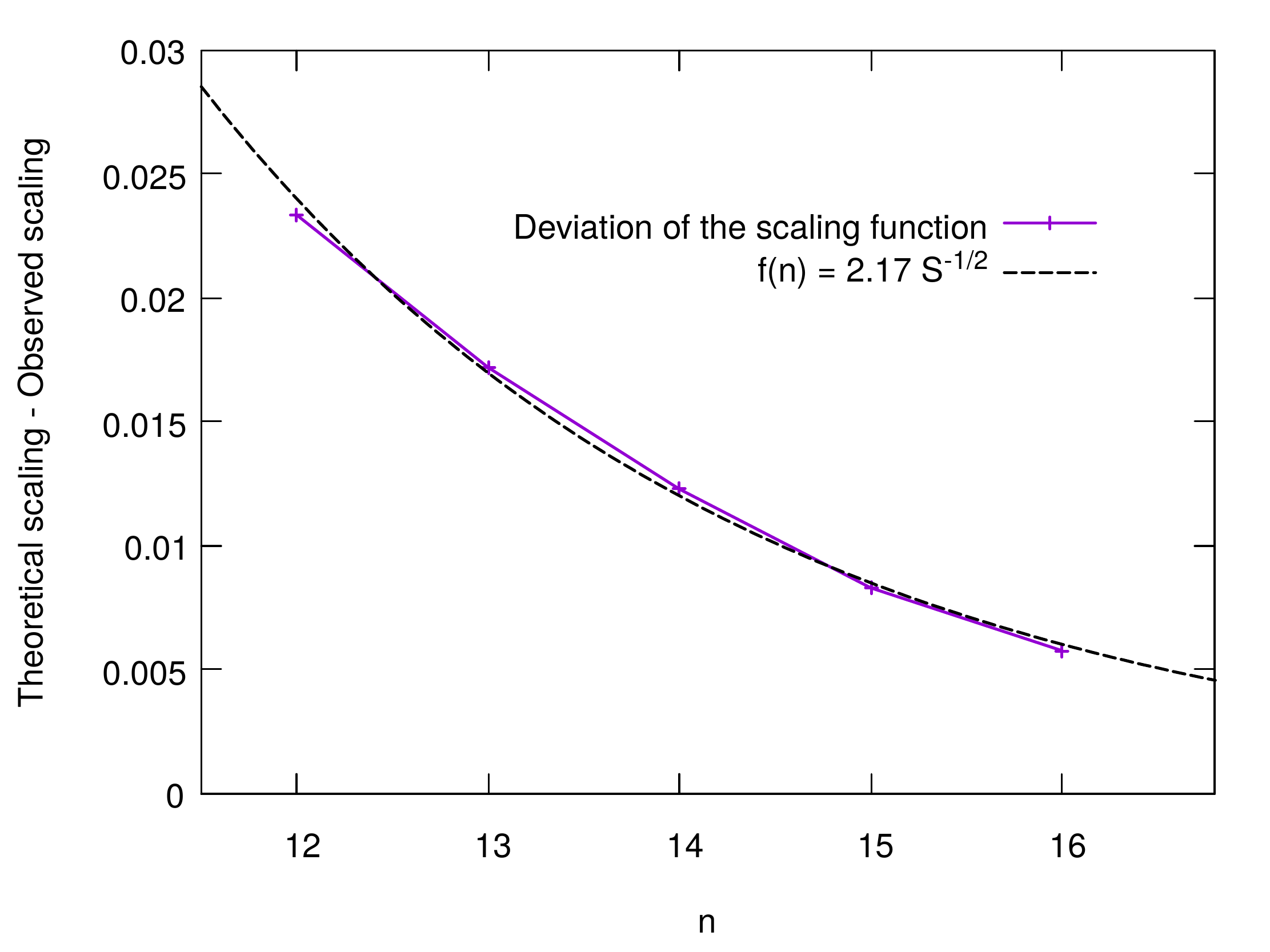}
    \caption{Difference between the observed scaling function and the theoretical (asymptotic) scaling function of the magnetic susceptibility at $\varepsilon S^{1/2} = -10$ as a function of n (here the number of sites per layer is taken to be $2^{n}$ and so, $S = 2.2^{n}$). The dashed line is the function proportional to $S^{-1/2}$.}
    \label{fig:corrections_scaling}
\end{figure}

From this, we expect the deviations from the asymptotic scaling function to be of the form $S^{-1/2}$. We verify this using our Monte-Carlo data by plotting the difference between value of the observed scaling function and the theoretically expected scaling function at a fixed value of $\varepsilon S^{1/2}$ (say $\varepsilon S^{1/2} = -10$) for lattices of different sizes. From Fig. \ref{fig:corrections_scaling} we see that the deviations from the theoretical scaling function do indeed vary as $S^{-1/2}$ which is consistent with our expectation.
%
\section{Summary and Conclusions}\label{sec:summary}  

In this paper, we studied the finite-size scaling corrections in the Ising model on random regular graphs of coordination number 3. We find that the Monte Carlo simulations data for the finite size scaling functions fits very well with the exact functional form we have calculated using the scaling theory, taking into account the dangerous irrelevant variable corresponding to quartic coupling. The corrections to scaling were also studied, and we verified that the deviations from the asymptotic scaling function falls off as $S^{-1/2}$. Lastly, we verified that the form of the scaling functions is the same for random regular graphs of coordination number 4 (not reported here).

\ack
The authors thank Prof.\;Kedar Damle for useful discussions. Suman acknowledges the National Supercomputing Mission (Param Bramha, IISER Pune) for the computational resources provided.
\vspace*{0.5 cm}  
\references


\begin{thebibliography}{99}       

\bibitem{Bethe1935} H. A. Bethe 1935 Proc. Roy. Soc. A {\bf 150} 552

\bibitem{Rushbrooke1938} G. S. Rushbrooke 1938 Proc. Roy. Soc. A {\bf 166} 296 


\bibitem{katsura}  S. Katsura and M. Takizawa, Prog. Theo. Phys., 1974  {\bf 51} 82. 

\bibitem{peruggi} F. Peruggi, J Phys. A: Math. Gen. 1983 {\bf 16} L713. 

\bibitem{WH1985} O Entin-Wohlman and C Hartzstein 1985 J. Phys. A: Math. Gen. {\bf 18} 315

\bibitem{BL1982} D. R. Bowman and K. Levin 1982 Phys. Rev. B {\bf 25} 3438(R)

\bibitem{ddhar_rltl} D. Dhar, R. Rajesh and J. Stilck 2011 Phys. Rev. E. {\bf 84} 011140.

\bibitem{cluster} A. Pelizzola, J Phys. A: Math. Gen. 2005 {\bf 38} R309-R339.

\bibitem{yedidia} J. S. Yedidia, W. T. Freeman and Y. Weiss, 2001 in Advances in Neural Information Processing Systems 13, eds. T. K. Leen, T. G. Dietterich, and V. Tresp, MIT Press 2001.
\bibitem{Kurata} M. Kurata, R. Kikuchi, and  T. Watari 1953 J. Chem. Phys. {\bf21} 434 

\bibitem{Rushbrooke} G. S. Rushbrooke and H. I. Scoins 1955 Proc. Roy. Soc. A {\bf 230} 74

\bibitem{PY1} J. K. Percus and G. J. Yevick 1958 Phys. Rev. {\bf 110} 1

\bibitem{PY2} M. Adda-Bedia, E. Katzav, and D. Vella 2008 J. Chem. Phys. {\bf 128} 184508

\bibitem{CT1_muller_zittartz} E. M\"uller-Hartmann and J. Zittartz 1974 Phys. Rev. Lett. {\bf 4}

\bibitem{CT2_Eggarter} T. P. Eggarter 1974 Phys. Rev. B {\bf 9}

\bibitem{CT3_Matsuda} H. Matsuda 1974 Progress of Theoretical Physics {\bf 51}


\bibitem{Baxter} R. Baxter 1982 \emph{Exactly Solved Models in Statistical Physics} (Dover Publications)

\bibitem{rgraphs_BA1} C. Baillee, D. A. Johnston, and J. P. Kownacki 1994 Nucl. Phys. B {\bf 432} 557

\bibitem{rgraphs_BA2} D. Dhar, P. Shukla, and J. P. Sethna 1997 J. Phys. A {\bf 30} 5259

\bibitem{rgraphs_BA3} A. Dembo and A. Montanari 2010 Annals of Appl. Prob. {\bf 20} 565

\bibitem{rgraphs_BA4} A. Dembo and A. Montanari 2010 Brazilian J. of Prob. and Stat. {\bf 24} 137

\bibitem{rgraphs_BA5}  D. A. Johnston and P. Plech 1997 Journal of Phys. A: Mathematical and General {\bf 30}

\bibitem{fisher1} M. E. Fisher, in Renormalization Group in Critical Phenomena and Quantum Field Theory, edited by J.D. Gunton and M. S. Green (Temple University, Philadelphia, 1974), p. 65.

\bibitem{privman1}  V. Privman (Ed.),  {\it Finite size scaling and numerical simulation of statistical systems}, 1990, World Scientific, Singapore.

\bibitem{cardy1} J. Cardy,  {\it Finite-size scaling}, 2012 Current Physics- Sources and Comments, North Holland, Amsterdam. 

\bibitem{brezin1} E. Brezin and J. Zinn-Justin, Nucl. Phys. {\bf B 257} [FS14] (1985) 867-893,  reprinted in \cite{cardy1}. 

\bibitem{rudnick1} J. Rudnick, H. Guo  and D. Jasnow, J. Stat. Phys. {\bf 41}  (1985) 353; reprinted in \cite{cardy1}.


\bibitem{binder1} K. Binder, Z. Phys. {\bf B 43} (1981) 119-140; reprinted in \cite{cardy1}. 


\bibitem{karsai} Marton Karsai, Robert Juhasz, and Ferenc Igloi, 2006,  Phys.  Rev.  E 73, 036116.


\bibitem{diam1} P. Erd\H{o}s and A. R{\'{e}}nyi 1959 Publ. Math. Debrecen {\bf 6}

\bibitem{diam2} B. Bollobas and W. Vega 1982 Combinatorica {\bf 2}


\bibitem{watts} D. Watts and S. Strogatz, Collective dynamics of ‘small-world’ networks. Nature {\bf 393}, 440–442 (1998).


\bibitem{wolff1} U. Wolff 1989 Phys. Rev. Lett. {\bf 62} 361--364. 

\bibitem{landau1} L. D. Landau 1937 Zh. Eksp. Teor. Fiz. {\bf 7}

\bibitem{landau2} L. D. Landau and E. M. Lifshitz 2013 \emph{Statistical Physics} (Elsevier)

\bibitem{abramowitz} M. Abramowitz 1974 \emph{Handbook of Mathematical Functions, with Formulas, Graphs, and Mathematical Tables} (Dover Publications)


\end{thebibliography}
\end{document}